\documentclass[%
 reprint,
 amsmath,amssymb,
 aps,
 prl,
]{revtex4-1}
\usepackage{graphicx}
\usepackage{dcolumn}
\usepackage{bm}

\begin{document}

\preprint{APS/123-QED}

\title{A Mean Field Platform for Excited State Quantum Chemistry}

\author{Jacqueline A. R. Shea$^1$}

\author{Eric Neuscamman$^{1,2,}$}%
\email{eneuscamman@berkeley.edu}

\affiliation{
${}^1$Department of Chemistry, University of California, Berkeley, CA, 94720, USA \\
${}^2$Chemical Sciences Division, Lawrence Berkeley National Laboratory, Berkeley, CA, 94720, USA
}

\date{\today}

\begin{abstract}
We present a mean field theory for excited states that is broadly analogous to ground state Hartree-Fock theory.
Like Hartree-Fock, our approach is deterministic, state-specific, applies a variational principle to a minimally correlated
ansatz, produces energy stationary points, relaxes the orbital basis, has a Fock-build cost-scaling, and can serve as the foundation for correlation methods such as perturbation theory and coupled cluster theory.
To emphasize this last point, we pair our mean field approach with an excited state analogue of second order M{\o}ller-Plesset theory and demonstrate that in water, formaldehyde, neon, and stretched lithium fluoride, the resulting accuracy far exceeds that of
configuration interaction singles and rivals that of equation of motion coupled cluster.
\end{abstract}

\maketitle

In a nutshell, Hartree-Fock (HF) theory
\cite{Szabo-Ostland}
applies the ground state variational principle to a wave function ansatz that includes only those correlations that are absolutely necessary to produce a qualitatively correct description of the electrons in a simple molecular ground state.
Indeed, HF theory's Slater determinant hews as closely as possible to a classical mean field state, in which the particles would be completely uncorrelated, while accommodating the Pauli correlations that must be included when describing electrons.
This simplicity keeps HF theory (relatively) affordable as it proceeds to minimize its variational principle and make the energy stationary with respect to changes in the molecular orbital basis.
That HF theory has for decades been the central platform on which high-accuracy weak correlation treatments are built
\cite{Bartlett:2007:cc_rev,Helgaker_book}
is a reminder of how valuable a qualitatively correct, minimally correlated wave function with relaxed orbitals can be.
For strongly correlated ground states in which an unambiguous minimal active space exists, the story is much the same,
but with complete active space self consistent field
(CASSCF) \cite{RUEDENBERG:1982:casscf,Werner:1985_1:mcscf,Werner:1985_2:mcscf,ROOS:1987:casscf}
and its descendants \cite{olsen1988ras,Malmqvist:2008p3011,Ghosh2008,Zgid2008dmrgscf,Gagliardi2013splitgas,Booth:2015:mcscf_fciqmc,Alavi:2016:fciqmc_casscf,Sharma2017sCICASSCF}
replacing HF theory as the basic platform upon which weak correlation methods are built.
\cite{LANGOFF:1974:davidson_correction,WERNER:1982:mrci,ANDERSSON:1990:caspt2,ANDERSSON:1992:caspt2,GDANITZ:1988:acpf,WERNER:1990:acpf,SZALAY:1993:mraqcc,ANGELI:2001:nevpt2,ANGELI:2002:nevpt2,ANGELI:2006:nevpt2,Neuscamman:2010:ct_review,Gagliardi2014,sharma2015mrlcc,Shiozaki:2015:NucGrad,Mazziotti2008casrdm}
In this paper, we develop an affordable mean field platform for simple excited states while also providing an initial weak correlation treatment and a discussion of how the approach can be generalized to strongly correlated excitations.

Unlike the situation for ground states, even zeroth order descriptions of most excited states require more than Pauli correlations.
For example, consider the open shell singlet of a simple HOMO$\rightarrow$LUMO excitation.
In this state, two opposite-spin electrons are strongly correlated with each other so as not to occupy the same orbital at the same time,
implying that a minimally correlated excited state ansatz must incorporate correlations not present in a Slater determinant.
An obvious candidate for this job is a configuration interaction singles (CIS) ansatz
\cite{HeadGordon:2005:tddft_cis}
which, to stay in line with ground state mean field methods like HF and CASSCF, has had its energy made stationary with respect to relaxations of the molecular orbital basis. 
While this direction has of course been explored before, previous approaches have tended
to approximate the orbital relaxation in order to suppress cost.
\cite{Liu2012,Liu2013,liu2014voacis,Veldkamp2015}
For example, Subotnik's OO-CIS approach \cite{Liu2012} achieves its efficiency by employing an
incomplete and HF-approximated Newton-Raphson optimization.
Inspired by the need for fully-relaxed excited state orbitals in situations like charge transfer and core spectroscopy,
we seek here to achieve them at the same cost-scaling as CIS by combining recent progress in excited state variational principles
\cite{Zhao:2016:dir_tar,robinson2017varmatch,Shea:2017:scesvp,ye2017sigma}
with a compound application of automatic differentiation.
To demonstrate the efficacy of this approach as a mean field platform on which to build excited state correlation methods,
we use it as the basis for an excited state analogue of second order M{\o}ller-Plesset (MP2) theory. \cite{Helgaker_book}

As we are looking to describe an excited state, minimizing the energy 
\begin{align}
E = \frac{\langle\Psi|\hat{H}|\Psi\rangle}{\langle\Psi|\Psi\rangle}
\end{align}
in pursuit of the ground state variational principle will not guarantee convergence to the desired state.
Instead, we will seek to minimize the Lagrangian
\begin{align}
\label{eqn:lag}
L = W + \vec{\mu} \cdot \frac{\partial E}{\partial \vec{\nu}}
\end{align}
in which the excited state variational principle \cite{Messmer:CPL:1970}
\begin{align}
\label{eqn:varprin}
W = \frac{\langle\Psi|(\omega - \hat{H})^2|\Psi\rangle}{\langle\Psi|\Psi\rangle}
\end{align}
guarantees that a sufficiently flexible wave function will converge to the exact (excited) energy eigenstate with energy closest to $\omega$ while the Lagrange multipliers $\vec{\mu}$ guarantee that an energy stationary point with respect to the wave function variables $\vec{\nu}$ will be achieved even when working with an approximate ansatz.
While we could have done without $W$ if we merely wished to ensure an energy stationary point was reached, its presence guarantees that in the limit of a sufficiently flexible wave function, the global minimum of $L$ will be the stationary point corresponding to the desired excited state. 
Although we do not pursue it here, one can imagine augmenting $W$ with functions of the dipole moment or other observables in order to differentiate between states that are energetically degenerate.


To make the minimization of this Lagrangian affordable, we must deal both with the difficulty of the $\hat{H}^2$ term and with the fact that derivatives of $L$ with respect to the variational parameters $\vec{\nu}$ (e.g.\ the CI coefficients and the orbital rotation variables) lead to second derivatives of the energy.
To avoid $\hat{H}^2$, we can resolve an identity in the basis of Slater determinants
\begin{align}
\label{eqn:wri}
W = \frac{1}{\langle\Psi|\Psi\rangle} \sum_{I}
\langle\Psi|(\omega - \hat{H})|I\rangle\langle I|(\omega - \hat{H})|\Psi\rangle
\end{align}
and, following the approach of coupled cluster (CC) theory, \cite{Bartlett:2007:cc_rev} make an approximation in which we restrict our attention (and the range of the sum) to the most chemically relevant corner of Hilbert space, in this case the span of the closed shell ``ground state'' determinant and the singles excitations.
Note that $W$ need not be evaluated exactly for the theory to operate correctly, as its role is merely to guide the optimization into the correct stationary point.
So long as $W$ provides a sufficiently strong nudge to get us close, the Lagrange multiplier term in Eq.\ (\ref{eqn:lag}) will ensure convergence to the stationary point.
This type of ``nudged'' convergence to a stationary point has been seen before in the application of full configuration interaction quantum Monte Carlo to excited states, \cite{Blunt:2015:excited_fciqmc} where a heavily approximated projection operator that could only serve to nudge the state propagation away from the lower states was sufficient to converge the imaginary time evolution onto excited states.
When an approximate nudge is insufficient, our formulation allows for a systematic retreat to safety via increasingly accurate evaluations of $W$, but we stress that in our initial testing this has yet to prove necessary.
Indeed, in all cases tested so far, the even more aggressive approximation 
\begin{align}
W \approx (\omega - E)^2
\label{eqn:wme2}
\end{align}
leads the minimization of $L$ to converge on to the same stationary point as when truncating
the sum in Eq.\ (\ref{eqn:wri}).
If this equivalence is maintained after more extensive testing in the future, there would be a strong
simplicity argument in favor of employing Eq.\ (\ref{eqn:wme2}).

Unlike the $\hat{H}^2$ terms, the challenge of second energy derivatives can be overcome by a compound
use of automatic differentiation (AD), \cite{Griewank-Walther-book}
which ensures that all of the derivatives of a many-input-single-output function can be evaluated
for a small constant multiple of the cost of evaluating the function itself.
As AD's use in quantum chemistry is still in its infancy,
\cite{sorella2010qmcforces,Neuscamman2013jagp,Aspuru-Guzik:2017:ad}
let us briefly explain the principles so that its usefulness for our purposes is clear.
First, consider that many complicated functions, such as the CIS energy, can be written as a
graph in which each node is one of the four basic binary operations shown in Table \ref{tab:ad}
in which two input quantities $a$ and $b$ go in and the output quantity $f$ that comes out may
then become one of the inputs for one or more of the other nodes in the graph.
One can compute the overall function value $g(\vec{x})$ by traversing the graph, starting from the dangling
edges that are the inputs $\vec{x}$ and moving forward through all the nodes until the final output $g(\vec{x})$ is reached.
Now, if one can afford to store the outputs of each node in the graph, it becomes possible to evaluate all of
$g$'s first derivatives with respect to the elements of $\vec{x}$ for a cost that is a small constant
multiple of the evaluation cost of $g$ via a sort of reverse traversal of the graph.
\cite{Griewank-Walther-book}
Crucially, if one considers a given node and assumes that they already know the partial derivative of
$g$ with respect to the output $f$ of that node, then the chain rule and
the simple derivative formulas in Table \ref{tab:ad} ensure that the partial derivatives of $g$
with respect to the node's inputs $a$ and $b$ can be evaluated via three or fewer binary arithmetic
operations.
Starting at the final node and working backwards, one finds that the number of operations required to get
all the derivatives of $g$ with respect to all the intermediates and all the elements of $\vec{x}$
is not worse than four times the number of nodes in the graph, and so the cost to get all the
derivatives $\partial g / \partial \vec{x}$ is a small constant multiple of the cost to evaluate $g$ itself.


\begin{table}[t]
\caption{Expressions for the derivatives of the four basic arithmetic functions
         in terms of the (presumably stored) values of their inputs $a$ and $b$
         and output $f$.
         \label{tab:ad}
}
\begin{tabular}{c r r r r}
\hline\hline
\vphantom{\Big(\Big)}
$f(a,b)$ & \hspace{4mm} $a+b$ & \hspace{4mm} $a-b$ & \hspace{4mm} $ab$ & \hspace{4mm} $a/b$ \\[2pt]
\hline
\vphantom{\bigg(\bigg)}
$\partial f/\partial a$ & $1$ \hspace{1mm} & $1$  \hspace{1mm} & $b$ \hspace{0.3mm} & $1/b$  \\[4pt]
$\partial f/\partial b$ & $1$ \hspace{1mm} & $-1$ \hspace{1mm} & $a$ \hspace{0.3mm} & $-f/b$ \\[4pt]
\hline\hline
\end{tabular}
\end{table}

This approach, known as reverse accumulation, provides a straightforward if tedious recipe for constructing a low cost implementation of analytic derivatives, and we may for example apply it to the CIS energy $E$ to obtain an efficient function for the energy first derivatives that appear in Eq.\ (\ref{eqn:lag}).
Folding this logic over on itself, we recognize that thanks to the dot product of these efficient derivatives with $\vec{\mu}$, $L$ itself can be implemented as a many-input-single-output function of the variables $\vec{\nu}$ and $\vec{\mu}$ whose cost is a small constant multiple of that of $E$.  By a second, compound application of reverse accumulation we may thus arrive at an implementation that delivers analytic derivatives of $L$ with respect to both $\vec{\nu}$ and $\vec{\mu}$ for a cost that is also a small constant multiple of that of a CIS energy evaluation
and with an additional memory requirement for intermediate storage that scales only as the square of the system size.
Note that, in practice, setting up reverse accumulation is tedious work that can and has been automated in many software packages.  In the present study, we have leveraged the machine learning community's rapid progress in reverse accumulation software by employing the TensorFlow \cite{tensorflow} framework to evaluate our Lagrangian's derivative vector, the norm of which we minimize by a quasi-Newton approach. \cite{Nocedal:1980:lbfgs}
As the cost of a CIS energy evaluation scales as the cost of a Fock matrix build, so does evaluation of $L$ and the necessary derivatives, leading our excited state mean field approach to have the same cost scaling as CIS.

While we have motivated this Lagrangian approach with the prospects of relaxing the orbitals in CIS,
the logic supporting its construction is much more general, and indeed
$L$ can in principle be employed efficiently 
with any ansatz for which $E$ and a reasonable approximation to $W$ can be efficiently evaluated.
$L$ could, for example, be used as a more rigorous alternative to maximum overlap methods \cite{gilbert2008mom}
when optimizing a CASSCF wave function for an individual excited state.
Although we are quite curious about this possibility, we do not pursue strongly correlated excited state treatments
in this study.
Instead, we focus on delivering a fully orbital-relaxed CIS wave function and testing its ability to
act as a platform for excited state correlation treatments in the same way HF theory does for ground states.

To this end, we employ the CIS-like ansatz
\begin{align}
|\Phi\rangle = e^{\hat{X}} \left( c_0|0\rangle + \sum_{ia} c_{ia}|{}^a_i\rangle + \sum_{\bar{i}\bar{a}} c_{\bar{i}\bar{a}}|{}^{\bar{a}}_{\bar{i}}\rangle \right),
\label{eqn:wfn}
\end{align}
in which excitations are labeled by alpha ($i$,$a$) or beta ($\bar{i}$,$\bar{a}$) indices and the closed shell determinant $|0\rangle$
is included to help the orbital-relaxed excited state better maintain orthogonality to the RHF ground state.
The vector of variables $\vec{\nu}$ that we optimize via $L$ includes the coefficients $c_0$, $c_{ia}$, and $c_{\bar{i}\bar{a}}$, as
well as the elements of the matrix $X$ that defines the orbital rotation operator,
\begin{align}
\hat{X} = \sum_{p<q} X_{pq} \left( \hat{a}^+_p \hat{a}_q - \hat{a}^+_q \hat{a}_p \right),
\end{align}
which for the present study we constrain so as to keep the orbitals spin-restricted.
Although we do not explore the possibility in the present study, this approach could be generalized to work
with a CASSCF wave function by replacing the CIS expansion in Eq.\ (\ref{eqn:wfn}) with the CASSCF CI expansion.
Given the much greater size of such CI expansions, it may in that case be more effective to use $L$ only
for the orbital rotation optimization and to instead rely on modern CI solvers to keep the energy stationary
with respect to the CI coefficients. 

While the excited state mean field (ESMF) ansatz $|\Phi\rangle$ is more flexible than CIS and might therefore be expected to be
more accurate, one should remember that, due to the significant effect of weak correlation on energetics,
HF theory itself is quite poor quantitatively even when it is qualitatively a good zeroth order wave function.
By the same reasoning, total energies and energy differences from ESMF wave functions (which revert to RHF when $\omega$ targets the ground state) are not likely to be competitive in accuracy with methods that incorporate correlation effects, such as equation of motion coupled cluster singles and doubles (EOM-CCSD). \cite{Krylov:2008:eom_cc_review}

As in the ground state, we are interested in the ESMF wave function not as a destination in and of itself, but rather as
a reliable platform upon which to construct correlation treatments that can reasonably hope to achieve more quantitative accuracy.
Considering first the prospects for a coupled cluster theory, we note that due to the natural termination of the
Baker-Campbell-Hausdorff (BCH)
expansion in traditional, similarity-transformed CC singles and doubles (CCSD), \cite{Bartlett:2007:cc_rev}
the usual approach of projecting the CC eigenvalue equation
\begin{align}
( e^{-\hat{T}} \hat{H} e^{\hat{T}} - E ) |\Phi\rangle = 0
\end{align}
into the space of low order (in this case internally contracted) excitations from the reference
will, as in the ground state theory, lead to a polynomially complex system of equations for the cluster amplitudes.
However, due to the fact that bare triples excitations would be present within the internally contracted doubles,
the cost scaling of such an approach would, although still polynomial, be substantially higher than in ground state CCSD.
More enticing is the prospect of leveraging the fact that the potentially long-range nature of the excitation
and the orbital relaxations it induces should already be accounted for in the ESMF reference state,
allowing lower-cost local CC approaches
\cite{schutz:2001:lcc,subotnik:2006:lcc,yang2012osvlcc,riplinger2013pnocc}
to focus on what they do best: treating short-ranged weak correlation.

\begin{table*}[t]
\caption{Comparisons for singlet excitations.
         For EOM-CC(2,3) we report excitation energies in eV,
         with other methods' results reported as excitation energy errors
         in eV relative to EOM-CC(2,3) and summarized in terms of
         mean unsigned error (UE) and maximum UE.  
         For stretched LiF, transitions are relative to the closed
         shell ionic state and are labeled by the F $\rightarrow$ Li
         orbitals involved (the bond is aligned along the $z$ axis).
         For other cases, transitions are relative to the ground state.
         Traditional methods were evaluated with Molpro \cite{MOLPRO_paper}
         and QChem. \cite{shao:2015:qchem}
         \label{tab:excitations}
}
\begin{tabular}{r l l r @{.} l r @{.} l r @{.} l r @{.} l r @{.} l r @{.} l r @{.} l}
\hline\hline
\multicolumn{3}{ c }{ } &
\multicolumn{2}{ c }{ \hspace{0mm} EOM \hspace{0mm} } &
\multicolumn{2}{ c }{ } &
\multicolumn{2}{ c }{ } &
\multicolumn{2}{ c }{ } &
\multicolumn{2}{ c }{ } &
\multicolumn{2}{ c }{ } &
\multicolumn{2}{ c }{ \hspace{0mm} EOM \hspace{0mm} } \\
\multicolumn{3}{ c }{ \hspace{0mm} State \hspace{0mm} } &
\multicolumn{2}{ c }{ \hspace{0mm} CC(2,3) \hspace{0mm} } &
\multicolumn{2}{ c }{ \hspace{0mm} CIS \hspace{0mm} } &
\multicolumn{2}{ c }{ \hspace{0mm} OO-CIS \hspace{0mm} } &
\multicolumn{2}{ c }{ \hspace{0mm} CIS(D) \hspace{0mm} } &
\multicolumn{2}{ c }{ \hspace{0mm} ESMF \hspace{0mm} } &
\multicolumn{2}{ c }{ \hspace{0mm} ESMP2 \hspace{0mm} } &
\multicolumn{2}{ c }{ \hspace{0mm} CCSD \hspace{0mm} } \\
\hline
\\ [-1.5ex]
\multicolumn{16}{ c }{ \hspace{0mm} H$_2$O, cc-pVDZ, $r=0.9614$ \AA, $a=104.4^\circ$ \hspace{0mm} } \\ [1.5ex]
$n$ & $\rightarrow$ & $\sigma^*$      &   8&22 &   0&96 &  -0&35 &  -0&18 &  -0&74 &  -0&01 &  -0&08   \\
$n$ & $\rightarrow$ & $\pi^*$         &  10&25 &   0&70 &  -0&32 &  -0&11 &  -0&77 &   0&00 &  -0&06   \\
$\sigma$ & $\rightarrow$ & $\sigma^*$ &  10&86 &   0&94 &  -0&26 &  -0&17 &  -0&73 &  -0&05 &  -0&06   \\
$\sigma$ & $\rightarrow$ & $\pi^*$    &  12&93 &   0&65 &  -0&26 &  -0&10 &  -0&83 &  -0&03 &  -0&04   \\
$\pi$ & $\rightarrow$ & $\sigma^*$    &  14&82 &   0&18 &  -0&47 &  -0&06 &  -0&82 &  -0&04 &  -0&01   \\ [1.5ex]
\multicolumn{17}{ c }{ \hspace{0mm} CH$_2$O, cc-pVDZ, geometry: B3LYP/cc-pVTZ \hspace{0mm} } \\ [1ex]
%
$n$ & $\rightarrow$ & $\pi^*$     &   4&22 &   0&40 &  -0&14 &   0&01 &  -0&98 &  -0&09 &  -0&05   \\
$\pi$ & $\rightarrow$ & $\pi^*$   &  10&02 &   0&26 &  -0&30 &   0&46 &  -1&12 &  -0&06 &   0&12   \\
$n$ & $\rightarrow$ & $\sigma^*$  &   8&70 &   1&68 &   0&50 &  -0&46 &  -0&71 &   0&09 &  -0&08   \\
[1.5ex] 
\multicolumn{17}{ c }{ \hspace{0mm} LiF, cc-pVDZ, $r=8.0$\AA \hspace{0mm} } \\ [1ex]
%
2p$_{x}$ & $\rightarrow$ & 2s      & -2&68 &   1&49 &  -3&03 &  -1&44 &  -1&02 &   0&10 &  -0&22   \\ 
2p$_{z}$ & $\rightarrow$ & 2s      & -2&68 &   1&44 &  -1&24 &  -1&31 &  -1&02 &   0&10 &  -0&22   \\ 
2p$_{z}$ & $\rightarrow$ & 2p$_z$  & -0&84 &   1&48 &  -3&20 &  -1&48 &  -1&02 &   0&10 &  -0&22   \\  [1ex]
\multicolumn{17}{ c }{ \hspace{0mm} Ne, cc-pVTZ \hspace{0mm} } \\ [1.5ex]
2s & $\rightarrow$ & 3p    &  \hspace{1.5mm} 64&32 & 
                              \hspace{0.0mm}  2&66 & 
                              \hspace{2.7mm}  1&74 & 
                              \hspace{2.0mm}  0&41 & 
                              \hspace{2.0mm}  1&36 & 
                              \hspace{2.9mm}  0&34 & 
                              \hspace{1.5mm}  0&06   
\\ [1ex]
\hline
\\ [-1.5ex]
\multicolumn{3}{c}{Mean UE}    &    \multicolumn{2}{c}{ }   &   1&07 &   0&98 &    0&51 &  0&93 &   0&08 &     0&10   \\
\multicolumn{3}{c}{Max UE}     &    \multicolumn{2}{c}{ }   &   2&66 &   3&20 &    1&48 &  1&36 &   0&34 &     0&22   \\
\hline\hline
\end{tabular}
\end{table*}

As intriguing as CC methods may be, the principle of Occam's razor suggests that an analogue of the much simpler MP2 theory
would be a wiser starting point for investigations into post-ESMF correlation methods.
We can arrive at just such a method, which we will denote as ESMP2, by applying standard Rayleigh-Schr{\"o}dinger perturbation
theory to the zeroth order Hamiltonian
\begin{align}
\hat{H}_0 = \hat{R} ( \hat{F} - \hat{H} ) \hat{R} + \hat{P} \hat{H} \hat{P} + \hat{Q} \hat{F} \hat{Q}
\end{align}
where $\hat{F}$ is the Fock operator with respect to the ESMF one-body density matrix, $\hat{R}=|\Phi\rangle\langle\Phi|$, $\hat{P}$ is the projector to the span of $|0\rangle$ and the singles excitations in the ESMF orbital basis, and $\hat{Q}=1-\hat{P}$.
Note that $\hat{P}|\Phi\rangle=|\Phi\rangle$ and $\hat{Q}|\Phi\rangle=0$.
This choice of $\hat{H}_0$ leads to the MP2-like zeroth order relationship
\begin{align}
\hat{H}_0 |\Phi\rangle = E_0 |\Phi\rangle = \langle\Phi|\hat{F}|\Phi\rangle |\Phi\rangle 
\end{align}
and indeed we see that when we set $\omega$ so as to target the ground state,
$|\Phi\rangle$ becomes the RHF state and ESMP2 simplifies to MP2.
In the excited state case, as in traditional MP2, the first order wave function
contains no determinants with fewer than two excitations, which is a consequence of including the $\hat{P} \hat{H} \hat{P}$
term in $\hat{H}_0$.
We also maintain the relationship $E_{\mathrm{ESMF}}=E_0+E_1$ in direct analogy to the
MP2 relationship $E_{\mathrm{HF}}=E_0+E_1$.
As the Fock operator is not diagonal for excited states, the first order amplitudes on doubles and triples 
are found by inverting $\hat{F}-E_0$ via the minimal residual Krylov subspace method,
for which the MP2-style denominators are an excellent preconditioner.
Note that this approach produces a fully excited-state-specific first order wave function, as opposed to the CIS(D) method
where the triples are a product of the CIS coefficients and the ground state MP2 amplitudes.
\cite{head-gordon:1994:cis-d,rhee2007scaled}

While we have, for the sake of simplifying development and testing, written our pilot ESMP2 implementation in a fully uncontracted form
whose triples part has an $N^7$ cost scaling, multiple avenues exist for recovering the $N^5$ scaling
of traditional MP2.
On the one hand, we could exploit sparsity in the ESMF coefficient matrix, which we observe
to have only a handful of elements that are not small.
Indeed, we have tested this idea by setting all but three of the elements to zero when solving the ESMP2
linear equation and evaluating its second order energy for the molecules discussed below and found that
excitation energy predictions are not strongly affected.
On the other hand, we could follow the internally contracted approach of Evangelista and coworkers.
\cite{Evangelista:2017:dsrg_excitation}


Table \ref{tab:excitations} reveals that, like its HF cousin in the ground state,
the ESMF approximation does not confer quantitative accuracy.
This behavior can be understood as a direct consequence of our design goal
of hewing as closely as possible to a classical mean field theory.
By including only those correlations
that are absolutely necessary to realize a fermionic excited state 
(namely Pauli correlations and the open shell correlation),
ESMF is missing all weak correlations and so, like HF, does not produce quantitative energies.
The fact that ESMF tends to underestimate excitation energies can also be understood in terms of what correlations are included.  
Indeed, in creating their open-shell arrangements, ESMF gives each excited state roughly one pair's worth of electron correlation, and so these states' energies are biased low compared to that of the closed shell ground state.
Thus, in direct analogy to HF theory, the simplicity and mean field nature of ESMF
prevents it from delivering accurate energetics on its own, but this of course
was not the intention.
What is more important is the question of whether ESMF can match HF theory's ability to
act as a foundation for correlation methods, a question that our early results
appear to answer strongly in the affirmative.

The data reveal that in water, formaldehyde, stretched lithium fluoride, and neon, ESMP2 rivals EOM-CCSD in accuracy.
Its errors are typically at least a factor of five smaller than CIS, and it substantially outperforms CIS(D)
in the charge transfer states of stretched LiF.
Our primary explanation for this success is that the reference wave function's mean-field
quality of having fully relaxed orbitals places it at a similar ``distance'' from the
correct wave function as for HF in the ground state, with the subtle
effects of weak electron correlation being all that is missing.
In contrast, CIS(D), which is also inspired by and closely entwined with MP2 theory,
is asking its perturbation to capture both correlation effects and orbital relaxation.
While the former are typically small in systems that are not strongly correlated,
the size of the latter is much more system dependent.
Table \ref{tab:excitations} shows that in the low lying transitions of water, in which
the overall spatial distribution of electrons is not greatly changed, both
CIS(D) and ESMP2 are highly accurate.
In stretched LiF, however, the transitions convert between ions and neutral atoms,
and as one would expect these large charge density changes lead to strong
orbital relaxation effects, as revealed by comparing the CIS and ESMF energies.  
These relaxations are much more difficult to treat perturbatively, and, to make
matters worse, cause the closed-shell-state MP2 amplitudes that all
CIS(D) states rely on to be less appropriate for the open-shell states.
Looked at from this perspective, it is not surprising that ESMP2, thanks to
its orbital-relaxed reference and fully state-specific perturbation,
delivers more uniform accuracy across
charge-transfer, valence, and Rydberg states alike.
Indeed, when compared to the high-level benchmark of EOM-CC(2,3), \cite{Hirata2000}
the maximum unsigned error of ESMP2 is less than the mean unsigned error for CIS(D).

We can also compare our results with more recent attempts to provide orbital relaxations for CIS.
The OO-CIS method, for example, provides relaxations via a single Newton-Raphson orbital optimization
step in which the HF Hessian is used as an approximation to the CIS orbital Hessian. \cite{Liu2012}
As OO-CIS lacks a treatment for weak correlation, it is not surprising that in H$_2$O, CH$_2$O, and Ne,
its accuracy is for most states better than CIS (which has neither orbital relaxation nor
a weak correlation treatment) but worse than CIS(D), EOM-CCSD, and ESMP2, as shown in Table \ref{tab:excitations}.
The large errors that OO-CIS makes in LiF can be understood as a consequence of its Hessian
approximation, as LiF's smallest HF Hessian eigenvalues are about an order of magnitude smaller than
those in the other molecules, and so when this Hessian is inverted to get the OO-CIS orbital
relaxation, the resulting Newton-Raphson step is much too large.
In this system at least, it appears that the HF Hessian is not an effective surrogate for the
CIS Hessian.
Unlike OO-CIS and our approach, the variationally orbital-adapted CIS (VOA-CIS) method seeks to provide
orbital relaxations through specially chosen expansions of the configuration interaction space. \cite{liu2014voacis}
However, there is at present no single prescription for defining the expansion
(the developers explore at least eight possibilities in their initial paper \cite{liu2014voacis})
and so instead of making an extensive and hard-to-interpret direct comparison, we will
point out the developers' conclusion that ``most of the time, VOA-CIS closely follows CIS(D)''
and their data that shows that it is not unusual for VOA-CIS to be in error by between 0.5 eV and 1 eV
for single excitation energies. \cite{liu2014voacis}
In summary, our preliminary testing shows ESMP2, with its inclusion of both orbital relaxation and state-specific
correlation, to be closer in its behavior to EOM-CCSD than to previous attempts at
augmenting the CIS wave function.

We have presented an excited state mean field theory and investigated its potential
as a platform on which to build excited state correlation treatments.
Like HF theory, the method relies on making a minimally correlated wave function's
energy stationary with respect to orbital relaxations.
While HF does this for a Slater determinant, we do so for a CIS-like wave function in order to
accommodate the basic structure of simple excitations, and so the cost of our
mean field optimization has the same scaling with system size as CIS.
Unlike HF theory, our approach incorporates an excited state variational principle into
its Lagrangian so that ground and excited states are treated equally. 
In exploring the method's potential as a platform for correlation treatments, we have
constructed an excited state analogue of MP2 theory and found that, in initial tests, its accuracy
rivals that of EOM-CCSD.

Looking to the future, there are a number of important questions to consider about this
excited state mean field approach.
First, how will it fare when applied to larger and more complicated
charge transfer systems, core excitations, and Rydberg states?
Second, will its advantages be maintained when paired with more complex and
expensive methods such as coupled cluster and the complete active space self-consistent field?
Finally, can this minimally-correlated, excited-state-specific
wave function be usefully employed as a replacement for the Slater determinant in an excited-state-specific
generalization of the Kohn-Sham approximation?
We look forward to exploring these and other questions that will undoubtedly arise in the
context of excited state mean field theory.

We thank Joe Subotnik for many helpful discussions,
and we acknowledge funding through the Early Career Research Program
of the Office of Science, Office of Basic Energy Sciences,
the U.S. Department of Energy, grant No.\ {DE-SC0017869}.
Calculations were performed both on our own desktop computers and at
the National Energy Research Scientific Computing Center,
a DOE Office of Science User Facility supported by the Office of Science of the U.S. Department
of Energy under Contract No.\ {DE-AC02-05CH11231}.


%

\end{document}